\documentstyle[epsfig]{aipproc}
\newcommand{\epsfigure}[2]{\epsfig{file=#1,width=#2}}

\long\def\symbolfootnote[#1]#2{\begingroup%
\def\thefootnote{\fnsymbol{footnote}}\footnote[#1]{#2}\endgroup}

\begin{document}

\title{Do theoretical physicists care about \\ the protein-folding problem?}

\author{Jos\'e L. Alonso$^{1,2}$, Gregory A. Chass$^{1,2,3}$,
Imre G. Csizmadia$^{3,4,5}$, \\
Pablo Echenique$^{1,2}$
\symbolfootnote[2]{Correspondence to P. Echenique: {\tt pnique@unizar.es}}
and Alfonso Taranc\'on$^{1,2}$}
\address{$^1$Departamento de F\'{\i}sica Te\'orica, Facultad de Ciencias,
Universidad de Zaragoza, \\
Pedro Cerbuna 12, 50009, Zaragoza, Spain.\\
$^2$Instituto de Biocomputaci\'on y F\'{\i}sica de los Sistemas Complejos
(BIFI), \\
Edificio Cervantes, Corona de Arag\'on 42, 50009, Zaragoza, Spain. \\
$^3$Global Institute Of COmputational Molecular and Material Science
(GIOCOMMS), \\
158-2 Major St., Toronto, ON, M5S 2L2 Canada. \\
$^4$Department of Chemistry, University of Toronto,\\
80 St. George St., Toronto, ON, M5S 3H6 Canada. \\
$^5$Department of Medical Chemistry, University of Szeged, \\
8 D\'om t\'er, Szeged, H-6720, Hungary.
}

\begin{flushright}
{\small
Included in the book: {\it Meeting on Fundamental Physics 'Alberto Galindo'},\\
\'Alvarez-Estrada R. F. {\it et al.} (Ed.),\\ Madrid: Aula Documental, 2004.
}
\end{flushright}

\maketitle

\begin{abstract}

The prediction of the biologically active native conformation of a
protein is one of the fundamental challenges of structural
biology. This problem remains yet unsolved mainly due to three
factors: the partial knowledge of the effective free energy function
that governs the folding process, the enormous size of the
conformational space of a protein and, finally, the relatively small
differences of energy between conformations, in particular, between
the native one and the ones that make up the unfolded state.

Herein, we recall the importance of taking into account, in a detailed
manner, the many interactions involved in the protein--folding problem
(such as steric volume exclusion, Ramachandran forces, hydrogen bonds,
weakly polar interactions, coulombic energy or hydrophobic attraction)
and we propose a strategy to effectively construct a free energy
function that, including the effects of the solvent, could be
numerically tractable. We also describe the situation in which the
native conformation has different covalent constraints than the
unfolded state (such as disulfide linkages), and then the exact native
structure can not be reached using the original free energy function.
Finally, we discuss about the limits and the lacks from which suffer
the simple models that we, physicist, love so much.

\end{abstract}

\newpage

\section*{Prologue \\ {\small (Selected memories of J. L. Alonso)}}

I became aware of Alberto Galindo in April 1965, when Professor Ortiz
Fornaguera, who made a report on a scholarship project of mine for the
{\it Fundaci\'on Juan March}, told me about him. The general ideas of
my project fit in a good proportion, as Ortiz said in his report, to
Bohm's remodeling of quantum mechanics.

At that time, I thought that there was nothing more beautiful than
quantum mechanics and relativity, and I knew by heart Terradas and
Ortiz's book on the subject \cite{Alonso:Terradas1952}.  Now, I think
that it was, above all, my ignorance which made me prefer what I knew
to some extent. But, after all, a first love is the most passionate
one.

The thing is that, when Ortiz, for whom I felt a deep reverence,
advised me to approach Alberto as the person who could best put my
ideas in order, I thought it would be great if Alberto Galindo would
agree to be my Ph.D.  thesis director.

In October that same year, a couple of days after the Pilar fiestas, I
was in Alberto's room in Zaragoza University, with my project about
hidden variables under my arm. I was surprised when he immediately
agreed to be my thesis director. Later, I learned that he was dead
short of assistants, as in the {\it Junta de Energ\'{\i}a Nuclear}, in
Madrid, a Ph.D. course had been organized which would be attended by
the best students in Madrid, Zaragoza, Valencia and Barcelona. That
course remains unsurpassed, despite uncountable attempts at
reproduction, and was crucial for the later development of Elementary
Particle Physics in Spain. So, to my ego's abasement, it was not my
project that affirmed Alberto's acceptance to be my thesis
director. But even if Alberto thought that an assistant fallen from
the sky was a present from the Virgin of the Pilar, it was not a bad
start.

As usual, along his magisterial scientific career, Alberto worried at
that time about subjects apparently as far apart from each other as
the uniqueness of the position operator for relativistic systems
\cite{Alonso:Galindo1965a} and the coupling of internal and space-time
symmetries \cite{Alonso:Galindo1965b}, both with large conceptual and
mathematical content. This helped in assuring that he was not
displeased that I devoted part of my time to pondering on the
philosophical foundations of quantum mechanics, so closely related to
the subject of hidden variables. I hope that my use of the term {\it
philosophical} is understood in this context and does not rise any
untimely debate.

According to my notes, my incursion into the subject ended when I told
Alberto that pursuing such a goal meant to consider time as an
operator.  Too strong an assumption both for him and for myself.

In this short recall of my links with Alberto in the last near 40
years, I must leave out many things, such as the unforgettable time we
spent in Orsay.  I shall skip too the crucial role he played in the
creation of the {\it Grupo Interuniversitario de F\'{\i}sica Te\'orica
(GIFT)}. I will only call the attention on his efforts to keep High
Energy Physics at Zaragoza at a high level, first convincing \'Angel
Morales and Rafael N\'u\~nez Lagos to exile in Zaragoza, where, at
that time, Francisco Yndur\'ain still was a Professor, and later,
along the years, favouring the collaboration of members of his
department in Madrid with members of ours in Zaragoza. I presume that
Guillermo Garc\'{\i}a will speak at length about the Madrid-Zaragoza
connection.

In any relatively important matter in which our department has been
involved, Alberto has involved himself. He did it in 1997, when he took
sides, together with some of us, dissuading the Government of Arag\'on
from carrying on the project of Accelerator Assisted Fission. The taking of
sides charged momentarily its toll, but Alberto is not the type man
to support a project which he deems badly conceived, even if he
is bound by a deep affection to its promoters.

More recently, in 2001, members of our department and members of the
departments of Biochemistry and Molecular and Cellular Biology,
Applied Mathematics and Condensed Matter of Zaragoza University
started the Institute of Biocomputation and the Physics of Complex
Systems (BIFI).  From the first moment, not only did several members
of the Theoretical Physics Department of Madrid Complutense University
got involved, but Alberto himself supported us most decisively in the
institutions in Arag\'on region requiring his advice. At the time of
writing, BIFI has members in the Complutense, Carlos III, Extremadura,
Elche and Granada universities and in the Institute of Science of
Materials of Madrid (CSIC).

For some of us, broadening our scientific interest to biophysics is turning
out to be a challenging experience. Our contribution to this book is in the
realm of the protein-folding problem.

\begin{figure}[t]
\begin{center}
\epsfigure{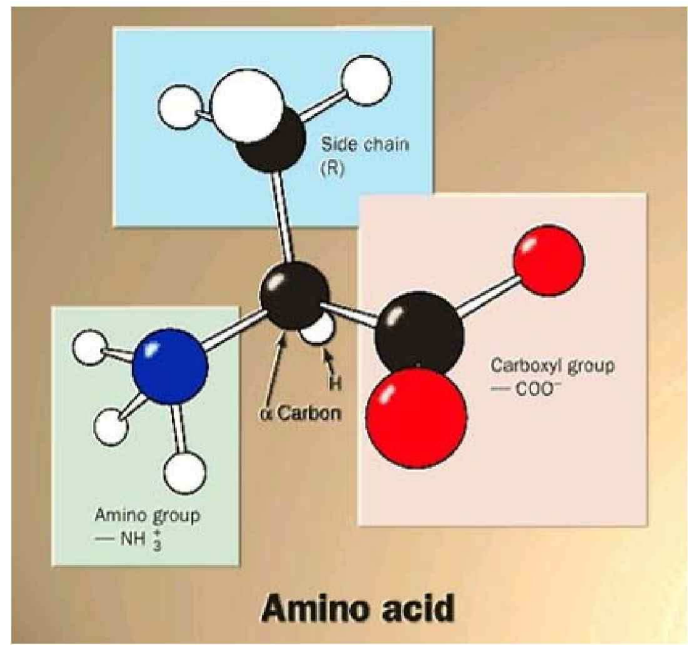}{9cm}
\end{center}
\caption{General schematic representation of an amino acid in its
zwitterionic form, the sidechain
part of the molecule is what distinguishes one amino acid from another. There
are 20 naturally occurring amino acids with sidechains of differing
complexity. The one depicted is one of the simplest chiral amino acids:
the alanine. }
\label{Alonso:Aacid}
\end{figure}

A protein is a biopolymer composed of monomer amino acidic {\it
building blocks} (See Figure~\ref{Alonso:Aacid}).  These biological
macromolecules fold at an amazingly fast rate, at least initially,
after being synthesized in the ribosome, and eventually reach a well
defined three-dimensional structure, known as {\it native
structure}. It is in this particular conformation that each protein
performs its specific biological task.  By unfolding and refolding the
ribonuclease protein {\it in vitro}, Anfinsen
\cite{Alonso:Anfinsen1973} showed that all the information needed to
reach the final native structure is encoded in the sequence of amino
acids, i.e., there is no cellular machinery needed to fold proteins
(which is true, at least, for most of them). What we now call the {\it
protein-folding problem} consists on the prediction of the native
structure of proteins starting from the knowledge of the amino acid
sequence and from the physical laws governing the interactions between
atoms in the frame of statistical mechanics.

\begin{quote}

I always feel like running away when any
one begins to talk about proteids in my 
presence. In my youth I had a desire to 
attack these dragons, but now I am afraid 
of them. They are unresolved problems 
of chemistry; and let me add, they are 
likely to remain such for generations to 
come. Yet every one who knows 
anything about chemistry and physiology, 
knows that these proteids must be 
understood, before we can hope to have a 
clear conception of the chemical processes 
of the human body \cite{Alonso:Ramsen1904}.

\begin{flushright}
      --  Professor Ramsen, Ira (1904)  --
\end{flushright}

\end{quote}

At the time when Professor Ramsen threw down the
gauntlet and challenged chemists to solving the
proteinfolding problem, he did not realize that 
the solution was to perhaps come from a myriad of
disciplines, not excluding those of theoretical and 
computational physics.

Our paper is organized as follows. In the next section we discuss
mesoscopic organization in biological and non-biological
matter. Afterwards, we describe the complexity of the proteins in
terms of their energy landscape and we talk about the Flory isolated
pair hypothesis in the context of these particular macromolecules. In
other two sections we recall what one can learn from simple models and
the necessity to go to detailed models if one aims at true predictive
power. The next section is devoted to the water inclusion, and in the
penultimate one we describe the two complementary approaches to the
protein-folding problem, the {\it holistic} and the {\it
reductionistic} one. Finally, in the last section, we discuss the
confluence of these two views in our goal of a step-by-step
construction of a detailed effective model with the final objective of
capturing the essential characteristics that make the protein-folding
process such an efficient and cooperative phenomenon.

\newpage

\section*{Introduction \\ {\small Are we talking physics?}}

In the year 2000, six {\it grand challenges} were identified and listed
in the final report of the {\it Physics in a New Era} series by a high-level
panel of the U.S. National Research Council (NRC), chaired by Thomas
Applequist of Yale University.

They are as follows:

\begin{enumerate}
\item Developing quantum technologies
\item Understanding complex systems
\item Applying physics to biology
\item Creating new materials
\item Exploring the Universe
\item Unifying the forces of Nature
\end{enumerate}

Regarding the challenge entitled ``Applying physics to biology'',
the mathematician Stanislaw Ulam once redefined the problem by
telling a biological physicist that he had finally understood the challenge:
``Ask not what physics can do for biology; ask what biology can do
for physics'' \cite{Alonso:Ulam2001}.

Possibly, Stanislaw Ulam meant that biology provides excellent
proving grounds for physics-based ideas about complex systems. But then
one would be addressing the second challenge, ``Understanding complex
systems'', rather than the third. One may then become suspicious that the
interest of physicists, at least of theoretical physicists, lies not in
biology itself but rather in ``as-yet-undiscovered organizing principles
that might be at work at the mesoscopic scales, intermediate between atomic
and macroscopic dimensions, and the implications of their discovery
for biology and physical science'' \cite{Alonso:Laughlin2000}. In the case
that the interest lies on mesoscopic organization, non-biological grounds like
spin glasses or strongly correlated electron systems are possibly easier
to code and more likely to allow new laws responsible for mesoscopic
organization to be extracted, if they exist at all. 

In this work, we shall approach a huge riddle, the high challenge for
science that is one of the most central phenomena of biological
processes: the protein-folding problem. We shall see that the uncertainties
involved in it are so overwhelming that, if one's purpose is to find
{\it new undiscovered organizing principles}, this
problem does not seem to be the best place to look, at least at the
present time. The unraveling of the mysteries of protein folding is, in
fact, a clear example of a problem that fits into the third grand
challenge of the NRC, ``Applying physics to biology''.

Then, if this third challenge does not interest theoretical physicists,
who is it addressed to? Perhaps to computational quantum chemists. Perhaps
it is such an interdisciplinary field that it places all disciplines
{\it at the brim of a nervous breakdown}.

Yet, we physicists are expert at modeling nature, and applying physics
to biology may mean a challenge to our modeling skills as impressive
as the discovery of new organizing principles. The most important ingredients
of a good model are its predictive power and its supplying us with a
perception of the reasons for that power, i.e., with a hint of {\it what is
responsible for what}. One might also add simplicity as an ingredient to good
models. It has certainly been, up to now, one of the most attractive features
of successful physical models. Now we know that the simplicity of the
Hamiltonians used in physics is due to the proximity to fixed points in
the parameter space of possible Hamiltonians. Behind that feature lies the
fact that the interesting behaviour of traditional physical systems does not
depend crucially on the details of the system. On the contrary, in complex
systems, such as proteins, the interesting behaviour does depend crucially
on the details.

If we perturb a protein {\it a little bit} (e.g. slightly altering the pH or
substituting just one selected amino acid in the chain), the folding process
may change dramatically and the biological activity of the protein may
cease altogether. The existence of allosteric proteins (which drastically
alter their shape and properties when they link a small regulating molecule
like a vitamin) is a good example of this fact. We are mindful of 
biological functions
associated with tiny structural details! Our very lives are at stake!

Even Professor Erwin Schr\"odinger himself would have agreed on this point.
In a set of lectures given in Dublin, in 1943
\cite{Alonso:Schrodinger1944}, he stated the following:

\begin{quote}

Every particular physiological process that we observe, either within
the cell or in its interaction with the cell environment, appears
---or appeared thirty years ago ---to involve such enormous numbers of
single atoms and single atomic processes that all the relevant laws of
physics and physical chemistry would be safeguarded even under the very
exacting demands of statistical physics in respect of large numbers [\ldots]
Today, we know that this opinion would have been a mistake. As we shall
presently see, incredibly small groups of atoms, much too small to display
exact statistical laws, do play a dominating role in the very orderly and
lawful events within a living organism. They have control of the observable
large-scale features which the organism acquires in the course of its
development, they determine important characteristics of its functioning;
and in all this very sharp and very strict biological laws are displayed.

\end{quote}

To sum it up, if we intend to model a protein's behaviour,
simplicity is best left aside and focus must be made on predictive power
and the capability to
enhance the perception of the reasons for that power. In contrast,
if the desire is to identify yet undiscovered organizing principles that
might be at work at mesoscopic scales, possibly there are other places to
look, perhaps in systems that do not suffer from so
much uncertainty.

\newpage

\section*{Principle of minimum frustration \\ {\small The correct free
energy function should help in solving the huge numerical problems}}

Today, we know that the two problematic paradoxes of protein science
are neither problematic nor paradoxical \cite{Alonso:Dill1999}

The first one of them is known as the {\it blind-watchmaker paradox}
\cite{Alonso:Dill1999,Alonso:Dawkins1986} and it equates the vastness
of the sequence space of polypeptides with the impossibility of ever
finding a protein-like sequence.

For a chain of, say, 100 natural amino acids, there are
$20^{100} \simeq 10^{130}$
possible sequences. Therefore, the probability of observing the
sequence of a particular protein is negligible. This problem has been regarded
as impossible to solve by creationists, who appeal to divine intervention
and has been circumvented by evolutionists through the mechanisms of
natural selection.

The fact is that there is not really a problem with the numbers. It has
been shown by statistical modeling \cite{Alonso:Chan1991} that the probability
of pulling out from a {\it soup} of random amino-acid sequences a particular
one that folds to the same structure and performs the same
biological function as protein $A$ is much more than
$10^{-130}$ (which would be the probability of extracting {\it exactly}
the sequence of $A$). It turns out to be more of the order of
$10^{-20}-10^{-10}$ \cite{Alonso:Dill1999},
which is a probability that, in spite of being still
small, can be easily overcome by natural selection. Besides, the probability
of pulling out a sequence that folds to any well defined native structure,
not just that of $A$, is even larger.

So it seems that there is an enormous degeneracy in sequence space, i.e.,
differences in sequence do not necessarily imply differences in 
biological function or
in fold. But, how can one easily explain this degeneracy? It is known that
the particular three-dimensional structure of a protein is determined to
a great extent by sidechain forces. In the folding process, these sidechain
forces cooperate with backbone forces to reach the stable native conformation
in $10^{-5}-10$ seconds after the sequence is synthesized at the ribosome.
Although the true balance between sidechain and backbone forces is not yet
known, it is impossible to fold a protein against the sidechain forces.
The backbone forces are essentially present with
the same magnitude irrespectively of the particular sequence and
it is the more sequence-specific sidechain
forces that help direct the fold. 
However, simple exact models \cite{Alonso:Dill1995}
show that the precise information of the sequence is, most of the times,
redundant; it has been found that the fold is primarily determined by
the sequence written in a two-letter alphabet rather than in the natural
twenty-letter alphabet.
One can classify the amino acids into two categories
regarding their affinity for water: hydrophobic (H) and polar (P).
Using this code, it is found that,
if a certain sequence does fold,
the sequence obtained by interchanging one hydrophobic amino acid for
another hydrophobic amino acid (analogously for polar ones) will fold
with a very high probability to a very similar structure.
Thus, the essential features
of the full $20^{100} \simeq 10^{130}$ sequences space remain in the
smaller space of the sequences written in the HP alphabet, which contains
{\it only} $2^{100} \simeq 10^{30}$ elements. Moreover, experiments
\cite{Alonso:Matthews1993a} show that only about $\onethird$ of the residues
are crucial for folding,
specifically those that define the hydrophobic core of the
protein. Adding this fact to the preceding one, we see that the real
search for protein structure takes place in a space whose size is closer
to $2^{100/3} \simeq 10^{10}$ than to the overwhelming $10^{130}$
first proposed, the remaining 120 orders of magnitude are highly degenerated.

As it was stated, $10^{10}$ is still a huge number, but it is affordable
for natural selection to search in such a space. Natural selection is
a {\it blind watchmaker} \cite{Alonso:Dawkins1986} that selects,
among these $10^{10}$
potential protein sequences, the true ones through a partially directed
process that, concurrently, involves considerable random choice
among alternatives.

Once Nature has selected a protein for its capability to fold and
to perform a certain biological function, one may wonder how this
molecule can find the native structure in such a short time, considering
the many degrees of freedom it has and the frustration among the different
interactions. This question was, for years, regarded as a paradox:
the {\it Levinthal paradox}.

It was first stated in a talk entitled ``How to fold graciously''
given by Cyrus Levinthal in 1969 and noted down by A. Rawitch
\cite{Alonso:Levinthal1969}. The paradox states that,
if in the course of folding
a protein is required to sample all possible conformations
(a hypothesis that ignores completely thermodynamics and statistical
mechanics) and the conformation of a given residue is independent of the
conformations of the rest (which is also false), then the protein will never
fold to its native structure.

For example, let us asume that each residue of a chain of 100 amino acids
can take up to 10 different conformations on average (typically, there are
9 relevant backbone conformations and a variable number of sidechain ones,
but let us use 10 for the sake of simplicity).
This makes a total of $10^{100}$ different conformations for the chain.
If the conformations were sampled in the shortest possible time
($\sim 10^{-13}$ s, i.e., the time required for a single molecular vibration),
one would need about $10^{77}$ years to sample all the conformational
space. This result implies that protein folding cannot be a completely
random trial-and-error process, as one already could have imagined by taking
into account the laws of thermodynamics and statistical mechanics.
Maybe, calling this problem a {\it paradox} is too much (in fact, Levinthal
did not use this word and, just after stating the problem, he explains
a possible solution to it).
It is clear that the size of the conformational space of a polypeptidic
chain is astronomically big (even if one takes into account the fact that
the conformations of different residues are not independent)
and we must explain how the system can
navigate through it from the unfolded state to the
native conformation in such short time.

Folding is not a general property of heteropolymers. Heteropolymers, due
to their many degrees of freedom and the many geometric constraints among
them, are said to present a great {\it frustration}, that is, 
there is not a single conformation of the chain which optimizes all
the interactions at the same time. In any conformation, the different
interactions are conflicting, i.e., {\it frustrated}. In polypeptides, as in
many heteropolymers, frustration is mainly due to chain connectivity
between monomers with opposite affinities to neighbours and/or environment.
This leads to a rugged energy landscape with many low-energy states,
high barriers, strong traps, etc.; up to a certain degree, a landscape
similar to that of spin glasses.
Hence, to reach a stable well-defined three-dimensional native structure,
a protein molecule cannot have a totally rough landscape. On average,
its native conformation must be more stabilizing than would be expected
for a polypeptide of random sequence \cite{Alonso:Hardin2002}.

\begin{figure}[t]
\begin{center}
\epsfigure{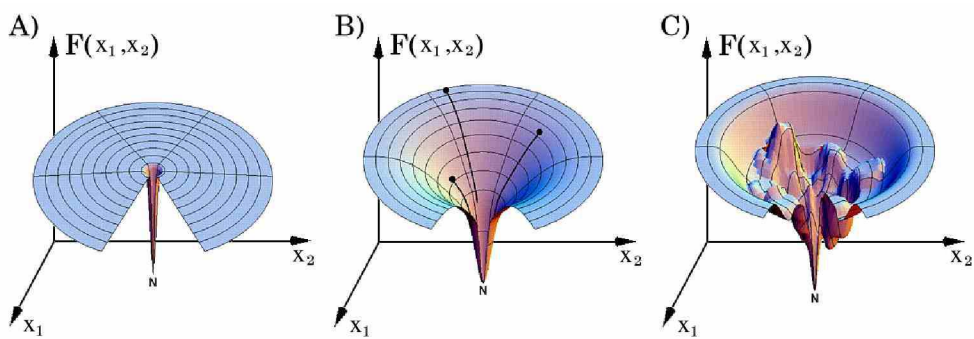}{13cm}
\end{center}
\caption{Possible energy landscapes of a protein. The conformational space
is asumed to be two-dimensional, the degrees of freedom being
$x_1$ and $x_2$. The free energy $F(x_{1},x_{2})$ is
a function of these two variables, which are internal degrees of
freedom of the molecule. The degrees of freedom of the solvent have
been {\it integrated out}. {\bf A)} The energy landscape as it would be
if the Levinthal Paradox was a real problem. {\bf B)} A smooth funneled
energy landscape. {\bf C)} A more realistic partially rugged funnel.
N stands for native state and it is assumed here to be the global
minimum.
(Figure taken from Dill K. A., Chan H. S., {\it Nat. Struct. Biol.}
{\bf 4}, 1 (1997) and somewhat modified) }
\label{Alonso:Funnels}
\end{figure}

Bryngelson and Wolynes \cite{Alonso:Bryngelson1981,Alonso:Bryngelson1989}
have termed this fewer conflicting interactions, than typically expected,
as the {\it principle of minimal frustration}.
This takes us to a natural definition
of a {\it protein} (opposed to a general {\it polypeptide}): a {\it protein}
is a polypeptidic chain whose sequence has been naturally selected to
satisfy the principle of minimal frustration. Such a molecule is allowed
to rapidly fold by trading conformational entropy for internal energy as it
moves down a landscape that is funneled (by virtue of the principle of
minimal frustration) towards the native structure
(see Figure~\ref{Alonso:Funnels}).

Recent theoretical and experimental evidence \cite{Alonso:Cheung2002}
fully support this folding scenario, i.e., biological proteins are only
minimally frustrated and the folded conformation
can be reached by one of a large
number of paths \cite{Alonso:Wolynes1995}.

In the context of protein folding, if one puts aside the explicit
solvent and agrees to treat it implicitly (which, given the present
power of computers and understanding of solvation at the molecular
level, is a must), the central physical object is the internal free
energy $F(\vec x)$ of the molecule (see
Figure~\ref{Alonso:Funnels}). This energy function depends only on
$\vec x$, the degrees of freedom of the polypeptidic chain.  The
degrees of freedom of the solvent molecules have been {\it integrated
out} and that is the reason of calling this energy {\it free}, meaning
that it depends parametrically on the temperature. It contains the
entropy of water, but not the conformational entropy of the
polypeptide (that is the reason of noting it as $F$ and not as $G$, we
reserve the letter $G$ for the total thermodynamic free energy of the
system, which only depends on the temperature and the pressure and not
on the internal degrees of freedom). Of course, if one works at
constant temperature, $F(\vec x)$ can be regarded as a normal
potential energy and a single conformation of the chain can be
considered as a unique microstate of the system in the mean field of
the solvent rather than a truly dynamical object. Finally, if one
refers specifically to the {\it graphical} features of $F(\vec x)$, it
is usual to use the term {\it energy landscape}.

We take into account that it is an experimental fact that the native
state consists of practically only one conformation or, more precisely,
a resonance of energetically-closely-spaced minima.
During the last quarter of the 20th century a hypothesis emerged, which
eventually became a dogma, namely, that
the native conformation must be the global
minimum of $F(\vec x)$. Some theoreticians agree about this point
\cite{Alonso:Anfinsen1973,Alonso:Anfinsen1975}, which in the context of
protein folding is referred to as the {\it thermodynamic hypothesis}.

\begin{figure}[t]
\begin{center}
\epsfigure{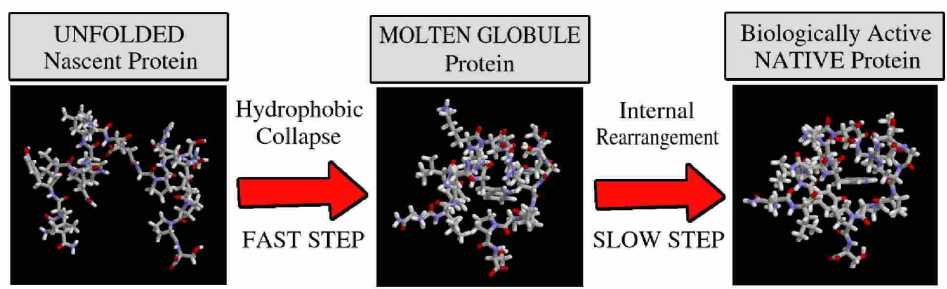}{13cm}
\end{center}
\caption{A schematic illustration of the multistep process of protein
folding as compressed into two major steps. Since the first step is
{\it fast} (a few miliseconds) it may correspond to a rolling down
process from the {\it highland area} of the potential free energy
hypersurface to the {\it lowland area}. Finding the final native
conformation on the {\it lowland area} is a considerably slower
process.}
\label{Alonso:MG}
\end{figure}

Currently, science has a better understanding of the mechanism of protein
folding: it consists of two complex steps interconnecting three phases
\cite{Alonso:Kim1990,Alonso:Matthews1993b,Alonso:Dobson1994,Alonso:Kuwajima1989,Alonso:Ptitsyn1991,Alonso:Csermely1999}, as shown schematically
in Figure~\ref{Alonso:MG}.

In the first step, the secondary-structural elements (e.g. helices,
$\beta$-pleated sheets, etc.) are formed and the hydrophobic
sidechains come into each other's vicinity, forming a hydrophobic
core. We will refer to this state, which has the majority of the
secondary-structural elements developed but whose tertiary structure
is not completely formed, as the {\it molten globule} state
\cite{Alonso:Ptitsyn1995,Alonso:Levitt1997}.

In the second step, the hydrophobic core is reorganized and a slow
search for the native structure begins. For some proteins, also
high-energy bonds are formed such as disulfide linkages (their
formation is \mbox{$\sim$ 8 $kcal \cdot mol^{-1}$} unfavourable) and
\mbox{{\it trans} $\leftrightarrow$ {\it cis}} isomerization of
proline peptide bonds occurs.  Formation of the disulfide linkages is
aided {\it in vivo} by {\it protein disulfide isomerases} and the cis
proline peptide bond isomerization is catalysed by {\it
peptidyl-proly-cis/trans isomerases}
\cite{Alonso:Raina1997,Alonso:Hamilton1998}.  When this happens, the
high-energy bond formations establish new constraints in the system,
favouring to populate a conformation that, in the former {\it
unconstrained} protein, would have been thermodynamically less
stable. This free-energy minimum of the new system (the protein with
redefined constraints) is the native state and it is very proximate in
the phase space to the conformations that made up the molten-globule
state. Of course, when no constraints redefinition occurs, the
stability of the native state is enough for the system to populate it
preferentially.

In the view of the aforementioned, it would seem reasonable to break
up protein-folding studies into {\it two steps}, along the lines
suggested by Figure~\ref{Alonso:MG}
\cite{Alonso:Ptitsyn1995,Alonso:Levitt1997}.  What is typically being
pursued with existing force fields, irrespectively whether by
Molecular Dynamics or by Monte Carlo search, would fit in the {\it
first step}, that is, in finding the conformations that make up the
molten globule state of the protein. The aim of {\it walking the last
part of the road} to the native state is still numerically unfeasible.
Methods must be designed in order to simulate this slow process or
a new generation of faster computers must be waited for. In the cases
in which the internal free-energy function $F$ does not include the
constraints present in the native state, i.e., the disulfide linkages
and the new value of the $\omega$ angles of some prolines, whose
description may be thought as a redefinition of the geometrical
constraints of the system, we need also to build new software
applications. {\it In vivo}, both the formation of the disulfide
linkage as well as the {\it trans} $\leftrightarrow$ {\it cis}
isomerization of the peptide bond, associated with the proline
nitrogen, are enzyme-catalysed. However, the process could be mimicked
without the involvement of the enzymes, as the thermodynamics of the
process is the same for both uncatalysed and catalysed reactions. To
the best of our knowledge, no systematic work has been carried out to
accomplish the construction of a force field aimed at simulating these
processes, but some preliminary research has been done in the field of
disulfide linkages formation \cite{Alonso:Viviani1998} as well as
\mbox{{\it trans} $\leftrightarrow$ {\it cis}} isomerization of
peptide bonds including both glycine and proline
\cite{Alonso:Baldoni2000,Alonso:Sahai2004}.

Under these conditions, a reasonable strategy to find the molten
globule, would be first to try to write a free energy potential
$F(\vec x)$ that incorporates, in the simplest possible way, all the
interactions that play an important part in the folding process.
Then, the problem is to locate the {\it lowland area} of $F(\vec x)$,
which depends on a large number of variables. Finding minima is a
classical problem of numerical analysis and a number of strategies are
possible (even {\it genetic} algorithms!). We will consider algorithms
closely related to statistical mechanics and based on Monte Carlo
simulations, which are efficient when a large number of degrees of
freedom are involved and present an important advantage: it is also
possible to simulate the behaviour of the system as a function of
temperature. To find minima using Monte Carlo methods, one must
replace the traditional approach (whose aim is the generation of a
Boltzmann ensemble of conformations) by an algorithm designed to
rapidly identify the lowland area of $F(\vec x)$
\cite{Alonso:Abagyan1999}. If this internal free energy function
properly describes the underlying physics, the giant size of the
conformational space is not a hindrance, as $F(\vec x)$ would be
funneled towards the molten-globule state and the efficiency of the
algorithm would be greatly increased.


\section*{Steric forces \\ {\small The importance of paying the fair
price (entropic)}}

To reach the more accurate internal free energy function mentioned
in the preceding section, it is desirable to find a set
of interactions such that, for the particular protein system and at
normal biological conditions (temperature, pH, salt concentration, etc.),
one can asume that factorization is possible, i.e., that the total energy
can be written as a sum of different components, one for each relevant
interaction. This possibility is dually convenient: on the one hand,
data to write each of the energy terms can be extracted from experimental
or {\it ab initio} results carried out on small simple systems in which only
a subset of the complete set of interactions is present and, on the other
hand, when testing the internal
free energy function, one can {\it switch on} or
{\it off} any of the interactions in order to gain insight about
{\it what is responsible for what}. However, factorization is not always
possible, because the different {\it interactions} in which one tries to
divide the problem are actually part of the same fundamental forces and, in
this way, a residual coupling may remain. Moreover, as we have already
mentioned, many microscopic degrees of freedom must be {\it integrated out}
if one wants to ever solve the problem. This elimination process causes
the remaining degrees of freedom to couple. The example of water is
illuminating: before the integration of the solvent degrees of freedom, the
electrostatic contributions as a function of atomic positions can be considered
practically pairwise; after the integration, however, the electrostatic
interaction energy of a pair of atoms depends on the positions of the rest.

In this section we shall talk first about one of the important
interactions that play a role in the folding of proteins, the steric
forces beyond nearest neighbours. We will dig into an example of how,
if one does not account in detail for these forces, the entropic price
that must be payed as the protein rolls down the real funneled
landscape becomes so distorted that there is no hope of ever getting
to a useful $F(\vec x)$. Obviously, the entropy of the unfolded state
is a determinant quantity in this discussion, because, the entropy of
the molten-globule state being very small, it is just the entropy of
the unfolded state that the favourable interactions must overcome in
order to fold the chain.

\begin{figure}[t]
\begin{center}
\epsfigure{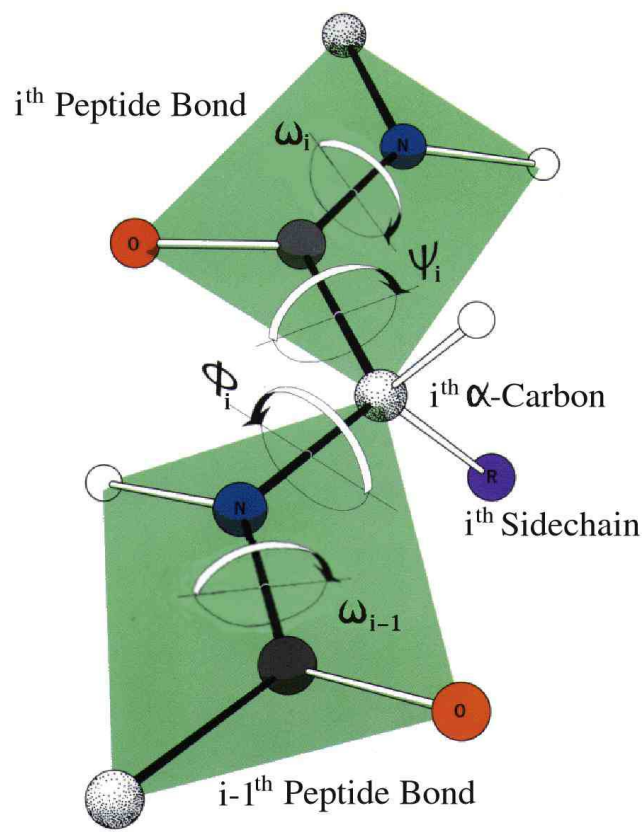}{9cm}
\end{center}
\caption{Angular degrees of freedom of the polypeptidic chain.
Ramachandran angles ${\Phi}_{i}$ and ${\Psi}_{i}$ are the most
flexible degrees of
freedom, the peptide bond angle ${\omega}_{i}$ is more rigid because of the
structure of the $\pi$ resonant bond.
(Figure taken from Voet D. and Voet J. G., {\it Biochemistry}, 2nd Ed.,
John Wiley \& Sons, 1995, and somewhat modified)}
\label{Alonso:Phipsi}
\end{figure}

Along many years, the original Zimm-Bragg \cite{Alonso:Zimm1959} and
Lifson-Roig \cite{Alonso:Lifson1961} helix-coil theories have been greatly
extended \cite{Alonso:Doig2002} with the aim of codifying experimental
data of peptide helices in solution \cite{Alonso:Munoz1994,Alonso:Munoz1997a}.
These theories have been markedly influenced by simplifying
{\it Flory's isolated-pair hypothesis} \cite{Alonso:Flory1969}, which states,
in the context of proteins, that each Ramachandran pair $({\Phi},{\Psi})$
\cite{Alonso:Ramachandran1963}
in the peptide backbone (See Figure~\ref{Alonso:Phipsi}) is independent.
It is rather obvious that this simplification suffer greatly from
over- or underestimations in order
to guarantee predictive power to the models that incorporate it, however
its utility in the codification of peptide helices data has been substantial
\cite{Alonso:Doig2002,Alonso:Munoz1994,Alonso:Munoz1997a}. In fact, it follows
from the hypothesis that {\it local} structure transitions are ruled out
as a possible source of cooperativity in protein folding
\cite{Alonso:Chan1995}; the entropic price is simply too high for short
polypeptide backbones to preferentially populate a small set of highly
similar conformations.

In a recent paper, Pappu, Srinivasan and Rose \cite{Alonso:Pappu2000} have
proved that, in the aforementioned helix-coil theories, the volume of
the coil region is much smaller than predicted by the isolated pair
hypothesis, thus lowering the entropic barrier for helix formation.
In a real protein, most conceivable states allowed by Flory's hypothesis
are not really accessible and steric effects beyond nearest neighbours
bias the unfolded molecule towards organized structure
\cite{Alonso:Srinivasan1999}. The {\it hard-sphere} model may be an
effective treatment of the most relevant steric repulsive interactions
beyond nearest neighbours \cite{Alonso:Pappu2000}.

Another important component of the free energy function $F(\vec x)$ is
the one that accounts for the backbone interactions between
neighbouring peptidic planes (See Figure~\ref{Alonso:Phipsi}). The
rotation around the Ramachandran angles $({\Phi},{\Psi})$ is hindered
by clashes between atoms around the $\alpha$-carbon and by strains in
the electronic clouds involved in the bonds. However, not all the
bonds in the backbone can rotate; the $\pi$ orbitals that partially
contribute to the peptidic bond severely limit the rotation around it
and that is the reason for the peptidic bond being nearly planar. This
fortunately allows us to decouple the potential energy surface (PES)
that describes the rotation around the Ramachandran bonds of one
particular residue from the PES of its neighbours, which has permitted
to render useful the {\it ab initio} quantum mechanical calculations
on dipeptides aimed to extract the one-residue PESs {\it in vacuo}
\cite{Alonso:Yu2001,Alonso:Perczel2003,Alonso:Vargas2002}.  This
established works show that the energy differences in the
$({\Phi},{\Psi})$ space can be very large: of the order of
\mbox{$\sim$ 10 $kcal \cdot mol^{-1}$}.  As this number is of the
order of the total folding energy, it is therefore clear that there
are zones of the Ramachandran space which are almost energetically
prohibited (these facts do not seem to change very much with the
inclusion of implicit solvent
\cite{Alonso:Iwaoka2002,Alonso:Topol2001}).  This component of the
free energy is crucial to be accurately implemented, as any error in
the relative energies between different values of $({\Phi},{\Psi})$
will be repeated as many times as the number of residues of the chain.

Two other relevant {\it in vacuo} interactions that remain to be
discussed are the coulombic forces (the portion of them that has not
been taken into account in the two preceding components) and the
hydrogen bonds. In fact, it is not clear whether hydrogen bonds must
be considered as being something essentially different to a
dipole-dipole interaction, however, it has been demonstrated that the
specific characteristics of these interactions, i.e. short range,
directionality and molecular orbital symmetry, are crucial to
stabilise important structures of the proteins, such as the
$\alpha$-helices \cite{Alonso:Pauling1951}.  Of course, when the
effect of water is added to this energy (a topic discussed further
on), some details of these interactions are modified. However, the
directionality and energetic influence of the hydrogen bonds between
peptidic planes remains practically intact, as their
quantum-mechanical {\it ab initio} characterization may be decoupled
from solvent effects \cite{Alonso:Ben-Tal1997}.

\section*{Simple models \\ \small What can we learn from them?}

As aforementioned, in molecular biophysics in general and in the
protein-folding problem in particular, it is not possible to design
accurate and precise simple models, that is, models that can
quantitatively predict at least some aspects of the behaviour of real
systems but, at the same time, contain a reasonably small number of
parameters.

Yet, the contributions of simple models have been qualitatively
relevant in the few last decades to shed light on the general
principles and mechanisms, as well as to seed new ideas concerning the
type of physics involved.  These contributions have allowed the ruling
out of some views that had been uncritically adopted from the domain
of small-molecule chemistry. The ideas of a folding funnel and of
parallel pathways in protein-folding dynamics stem from the results of
simple physical models, which also contribute in proposing possible
scenarios for molecular evolution, in describing the denaturation of
DNA, in predicting the secondary structure of RNA and in a number of
other domains of molecular biophysics
\cite{Alonso:Dill1995,Alonso:Galzitskaya1999,Alonso:Munoz1999,Alonso:Bruscolini2002,Alonso:Bruscolini2003,Alonso:Onuchic1995}.

Among the studies aimed at increasing understanding of the {\it common
characteristics} of globular proteins, those that deserve special
attention are the ones that show these characteristics arising from
considerations of symmetry and geometry associated with the
polypeptide chain and that only a limited number of folds arise from
geometrical constraints imposed by sterics and hydrogen bonds
\cite{Alonso:Hoang2004}.  Any more sophisticated model that wants to
take a step further in predicting the details of the folding of a
particular protein will not be able to contradict this type of
analysis. It must go further by incorporating details of the steric
forces, the hydrogen bonds and at the same time of long-range
coulombic interactions and others of entropic origin that, cooperating
with the former ones, stabilize one appropriate fold among those that
symmetry and geometry have previously selected.


\section*{Detailed models \\ \small Some of their ingredients and
difficulties}

A detailed model does not necessarily have to include all the atoms of
the molecule, nor does any one atom require treatment at the same
level of coarseness in all the interactions in which it plays a
role. For example, as mentioned, a hard-sphere description of an atom
may be appropriate to account for its repulsive steric interactions
with other atoms in the amino-acid chain, on the contrary, it is
completely inadequate to reproduce the PES associated with the
Ramachandran degrees of freedom. With such a description, all allowed
$({\Phi},{\Psi})$ pairs would have the same energy, a result that {\it
ab initio} studies
\cite{Alonso:Yu2001,Alonso:Perczel2003,Alonso:Vargas2002} prove
absolutely false.

Clearly a detailed model of any physical system need not to include
all the degrees of freedom, although it does have to include those
that are considered the largest contributors to the partition function
of the system or, alternatively, to an equivalent effective system, if
a part of the system (we are thinking about, for example, the part
that corresponds to the water molecules in a protein system) has not
explicitly been taken into account. Generally speaking, the
contribution to the partition function of the vibrational degrees of
freedom is negligible compared to that of the rotational ones at 300 K
(See table 11.2 in ref.~\cite{Alonso:Dill2003}), however, not all the
rotational degrees of freedom themselves play an equivalent role. In
the case of proteins, the dihedral angles around the $N-C_{\alpha}$
bond ($\Phi$) and around the $C_{\alpha}-C'$ bond ($\Psi$) determine
the {\it main-chain} or {\it backbone} geometry, from which stem the
different sidechains and must be undoubtly incorporated to any model,
as they define the secondary and tertiary structure. On the contrary,
the rotational degrees of freedom of the sidechains, which are
responsible for the different conformations of these groups, known as
{\it rotamers}, are certainly of secondary importance when compared to
the Ramachandran angles $({\Phi},{\Psi})$ and may be implemented in a
more simplified way.

Once the relevant degrees of freedom of the system are known, in order
to have any opportunity to find the molten globule state, we must
write a free energy function detailed enough. As it has been already
explained, the relevant interactions that must be included are: steric
constraints
\cite{Alonso:Flory1969,Alonso:Ramachandran1963,Alonso:Ramachandran1968,Alonso:Baldwin1999},
hydrogen bonds \cite{Alonso:Pauling1951,Alonso:Eisenberg2003},
Ramachandran PESs
\cite{Alonso:Yu2001,Alonso:Perczel2003,Alonso:Vargas2002}, coulombic
interactions and hydrophobicity \cite{Alonso:Kauzmann1959}.

On the topic of the different approaches to the protein-folding
problem, R. A. Abagyan \cite{Alonso:Abagyan1997} has classified them
in a way that is much liked by the authors of this article. According
to Abagyan, we can divide the scientists that try to predict the
native structure of the proteins into three categories: {\it
dynamicists}
\cite{Alonso:VanGusteren1977,Alonso:Brooks1983,Alonso:Mazur1991}, {\it
minimalists}
\cite{Alonso:Levitt1975,Alonso:Go1978,Alonso:Miyazawa1985,Alonso:Srinivasan1995,Alonso:Kolinski1994,Alonso:Rykunov1995}
and {\it synthesists}
\cite{Alonso:Anfinsen1975,Alonso:Abagyan1994,Alonso:Pedersen1995,Alonso:Auspurger1996}. In
Abagyan's own words:

\begin{quote}

Dynamicists believe that sufficiently long simulations of a
quasi-continuous trajectory of molecular dynamics of atomic models
{\it in vacuo} or in water will solve the problem using new
generations of computers, code parallelization, and optimized
simulation techniques.  Minimalists, unwilling to play power games and
too impatient to wait until new generations of processors cover the
next mile of a hundred mile road, simplify the system by using a
reduced atomic representation or a lattice, inventing a potential and
then enjoying the luxury of always finding the global minimum of their
energies as well as most of the other possible states for a chain of
up to a hundred simplified residues
\cite{Alonso:Kolinski1994,Alonso:Rykunov1995}. The third school shares
the impatience of minimalists, yet resists the temptation to use
simple models since it appears that accuracy is a pivotal
issue. Synthesists focus on the development of algorithms to replace
molecular dynamics as a generator of conformational changes and the
design of methods of energy calculation which combine accuracy and
speed.

\end{quote}

Abagyan includes himself in the group of the synthesists and gives
some more detailed characteristics of this approach. Since the
synthesists are, of the three categories, the scientists which we also
find ourselves in better agreement with, we quote again the words of
Abagyan to describe some of our shared ideas. Although points 1 and 3
require further qualifying.

\begin{quote}

Let us list some of the ideas and assumptions of the synthesists,
including the author, which the following review is based upon:

\begin{enumerate}
\item Oscillations of bonds lengths and bond angles are not essential
for protein structure prediction and some of these degrees of freedom
are not even excited at room temperature. Therefore, using torsion
angle space instead of Cartesian coordinate space is highly preferable
since it reduces the dimensionality of the problem by a factor of 7,
eliminates fast oscillations and smoothens the energy landscape.
\item A continous molecular dynamics trajectory is not really
necessary for structure prediction. The optimal structure can be found
by a global optimization algorithm making much larger steps.
\item Explicit consideration of water molecules can also be sacrificed
in simulations of folding for the following reasons: (i) too many
additional degrees of freedom; (ii) a really large box is necessary
because of the long-range nature of the electrostatic interactions;
and (iii) the relaxation times of water molecules after a large
conformational change is prohibitively long. Concurrently, the
solvation effect can be evaluated by continuous approximations more
efficiently and, potentially, more accurately.
\item The correct conformation and an enormous number of altenative
conformations of a polypeptide chain may have very close energies.  A
high accuracy of energy calculations is absolutely essential to
recognize the correct answer.
\end{enumerate}

\end{quote}

To conclude this section, an implicit {\it bona fide} assumption has
been held, namely, that one will be able to write a physical $F(\vec
x)$ that will produce a funneled landscape and that numerical
algorithms to find minima will smoothly roll down the funnel in an
acceptably short time. A more realistic scenario forces one to realize
that a free-energy function that is slightly different from the real
one combined with the high dimensionality of the phase space could
cause significant numerical problems. Even if there is no strong
frustration phenomena, to travel around (almost) all phase space is
very time consuming (ergodicity). If we do not study large regions, we
will not be able to guarantee that any lowland area found is the
lowest one. When a large number of degrees of freedom are considered
and one wants to study a true protein-folding problem, Monte Carlo
simulations are the most efficient algorithms to obtain
minima. However, continued improvement of these algorithms is needed
in order to speed up simulations and to facilitate more efficient
parallelization.


\section*{The water \\ \small Where the whole story takes place}

The actual shape of the free-energy fuction of the protein $F(\vec x)$ is
determined to a large extent by the fact that the folding takes place in a
particular environment: an aqueous one. Let us examine the basis for this.

Although the Ramachandran PESs and the steric volume exclusion have
their origins in strong {\it electronic} interactions, the weak
intramolecular bonds in the protein have a crucial role in the
preference of the differing conformations of the polypeptide chain.
The energy of formation of the weak bonds lies in the interval of
$1-5$ $kcal \cdot mol^{-1}$ (See the book by Dill and Bromberg
\cite{Alonso:Dill2003} for an inspired discussion about the hierarchy
of the energies of the different molecular interactions). At room
temperature, the average kinetic energy of a water molecule is
\mbox{$RT{\simeq}0.6$ $kcal \cdot mol^{-1}$}, providing the numerous
water molecules with sufficient energy to break these weak bonds.
Therefore, at physiological temperatures ($\sim 37^{o}C$), the weak
bonds have, on the one hand, a transitory existence, although their
joint action stabilize the relevant structures of the chain and, on
the other hand, they are strongly dependent on the solvent. Moreover,
the so called hydrophobic effect, which is an entropy-driven force, is
genuinely characteristic of the solvent.

It is clear that, for a certain interaction to be relevant in the
determination of the molten-globule state of a protein, it is a
necessary condition that it changes the order of the different
conformations on the energy axis determined by the rest of the
interactions.  The inclusion of water does change this order, even at
such a fundamental level as the one-residue Ramachandran PESs
\cite{Alonso:Iwaoka2002,Alonso:Topol2001} and hence why one is forced
to take it into account. Of course, an interaction that does not
change the energetic order of the conformations may still have an
effect if one makes a molecular dynamics simulation, as the rates at
which the transitions occur depend on the differences of energy and
not only on their order. However, if one wants only to minimize the
energy and to characterise the global minimum, although such an
interaction may affect the convergence rate of the algorithms used, it
will not change the fundamental features of the methods. The same
criteria may be applied to the approximations used to compute the
different interactions \cite{Alonso:Alonso}, i.e., one must check
whether or not the particular approximation changes the energetic
ordering of the conformations.

Regarding the question of how to implement the influence of water
\cite{Alonso:Perczel1995}, it seems reasonable to assume that the
effects due to the discreteness of the solvent are related only to the
first layer of molecules around the solute. This first shell of
molecules (the primary solvation layer) will be responsible for the
entropic forces associated with a perturbation of the hydrogen-bond
network and with the reduction of the phase space of water, which has
traditionally been termed the {\it hydrophobic effect}
\cite{Alonso:Tanford1979,Alonso:Sharp1991,Alonso:Chan2002}. To take
into account these forces and also those of enthalpic origin,
appearing when the primary solvation layer is confronted with a polar
atom, it is common practice to write an energy proportional to the
{\it Solvent Accessible Surface Area (SASA)}
\cite{Alonso:Eisenberg1986,Alonso:Lee1971} of each atom. Beyond the
first layer, water may be regarded as a linear isotropic and
homogeneous dielectric medium. Neither in one zone nor in the other
can one as yet afford the computational expense of treating water
explicitly at the molecular level. The increase in numerical
complexity would render the simulations near impossible at the present
time, as we have already remarked when quoting R. A. Abagyan on the
topic.

The rapid calculation of the energy of a conformation is a central issue
in protein folding, and the treatment of water remains a problem since
even the most approximated methods are relatively slow. To illustrate,
we mention the molecular dynamics simulations conducted by V. Pande
{\it et al.}
\cite{Alonso:Zagrovic2001}. These lengthy calculations have been performed
with the Jorgensen's optimized parameters for liquid simulations (OPLS)
force field \cite{Alonso:Jorgensen1988} and, to account for the solvent,
an approximated method
called GB/SA \cite{Alonso:Qiu1997}, which not only avoids using
explicit water but also approximates the Poisson equation for the dielectric
region. The inclusion of explicit water molecules would raise the
simulation times by approximately three or four orders of magnitude.

Pande's research group has used distributed computing techniques and a
{\it supercluster} of thousands of computer processors around the
world to address important questions regarding the $\beta$-hairpin
folding (a fragment of a G protein) of only 16 residues. Although the
results obtained from this massive simulation are very relevant, they
disagree with a proposed mechanism
\cite{Alonso:Munoz1997b,Alonso:Munoz1998} in which folding initiates
at the turn of the $\beta$-hairpin and propagates downwards in a {\it
zipper-like} way. Even if this disagreement is possibly due to the
model used \cite{Alonso:Munoz1997b,Alonso:Munoz1998}, we quote a
revealing comment in the article by Pande: ``The results of our study
generally speak in favour of using the OPLS potential set for studying
of protein fragments and small proteins. {\it However, the possibility
that this set is not perfectly suitable for the direct folding of the
$\beta$-hairpin remains}.''

This assertion shows that the design of a truly physical free-energy
function is a more fundamental, and certainly prior, issue than the
numerical problems associated with the simulations. This task is a
challenge to our modeling capacity that could be to the physicists'
liking.

\section*{The two sides of the coin \\
\small Holistic and reductionistic approaches to protein-folding}

There are two types of approaches in all scientific problem: {\it holistic}
and {\it reductionistic}. Clearly, the field of protein folding is no
exception! They consist, respectively, of the following:

\begin{enumerate}
\item Study the whole problem even if one must give up accuracy by
using a non-rigorous method.
\item Study a small fraction of the whole problem, using a rigorous method.
This is done in order to solve the whole problem more accurately.
\end{enumerate}

What we have discussed so far is mostly the holistic approach, in
which we take the whole problem and try to engineer a solution for it.

In contrast to that, there is the reductionistic approach, whereby one
studies peptide folding, rather than protein folding, using first
principle methods such as {\it ab initio} Hartree-Fock or Density
Functional Theory with Gaussian split-valence basis sets of different
sizes in order to solve the many electron Schr\"odinger equation. The
thesis is that any understanding we might obtain from peptides will
help us to understand protein folding. This has been discussed at some
length recently \cite{Alonso:Chass2001}.

Of course, the key question now is how long a polypeptide chain is
called a {\it peptide} and how long it is called a {\it protein} (or a
{\it polypeptide} if it does not fold to a well defined
three-dimensional structure).  Perhaps the border line between the two
fields lies around 30 residues in the chain.

The 1990s have seen a lot of development on this
\cite{Alonso:Perczel1991,Alonso:Perczel2000}.  Several
single-amino-acid diamides have been studied computationally and, by
the dawn of the 21st century, all naturally ocurring amino acids
(i.e. those with DNA codons) have been studied at least in an
exploratory way \cite{Alonso:Chass2002a}.  The full conformational
space of the dialanine dipeptide has been explored
\cite{Alonso:Perczel1993,Alonso:McAllister1993,Alonso:Perczel1994} and
certain motifs of tri- and tetrapeptides have also been studied in a
preliminary way. At that time, these achievements were possible by
pushing the computational limits. Today, even larger tetrapeptides can
be studied relatively easily \cite{Alonso:Borics2003}.

Before the turn of the millenium, HF/3-21G calculations were carried
out on a helix containing 12 alanine units. Only the 8 alanine unit
has been studied further, at the DFT level, using a larger basis set
and with solvation included \cite{Alonso:Topol2001}. We may hazard a
guess that, within a couple of years (i.e. by the years 2006-2008), a
helix containing 24 alanine residues will be computable at a similar
(or even more accurate) level of theory.  This particular length is
very relevant, as it is the length of the section of the transmembrane
proteins passing through the cellular wall
\cite{Alonso:Patel1999}. Soon thereafter, a 30 residue polypeptide
will be within the realm of the possibilities.  Such a long
polypeptide chain is no longer as flexible as the short peptides and
it already posseses some of the intrinsic properties of proteins.

As a reminder, we should note that so far nobody explored the full
conformational space of trialanine, which may have up to $9^{3}=729$
backbone conformers \cite{Alonso:Cheung1994,Alonso:Sahai2003a}.  A
tripeptide can offer an important structural feature, namely, that the
central residue has a nearest neighbour at both sides. Similarly,
tetrapeptides, such as tetraalanine, can offer a situation where a
$\beta$-turn has nearest neighbours at both ends, however, they may
exhibit up to $9^{4}=6561$ backbone conformations, which presently
represent a computational hindrance.

It is hoped that, when these territories are covered, involving a number
of different sequences of
amino acid residues, not only alanine, a deeper insight
will emerge and the computed data that support this insight will be useful
for those dedicated to the holistic approach.

\section*{What does the future hold? \\
\small Using high-level ab intio quantum molecular computations}

In the design of simple models, one must recognise that the {\it
future} always arrives sooner than one is prepared to give up the {\it
present}. The {\it present} becomes the {\it past} and challenges that
were put aside as being those of future work, soon require immediate
attention.

The protein-folding problem posesses a number of these
{\it future} challenges, including the treatment of explicit
solvating particles (particularly the primary solvation
layer), inclusion and evaluation of all thermodynamic
parameters, analysis and understanding of the bases of
molecular orbital symmetries and overlaps, in-depth 
systematic conformational analysis, as well as
required compromises in the thoroughness of the theoretical
treatment of the model peptide systems.

An ongoing reductionist (micro-molecular) {\it correction}
to more holistic (macro-molecular) models may be a
promising route to more accurate development of simple
protein models \cite{Alonso:Chass2001}.

One therefore requires an effective and analytical description
of the whole problem, including therein a numeric definition
of the relative spatial orientation of all constituent 
atomic nuclei \cite{Alonso:Chass2002a}.
Modular in design, of objective numerical
efficiency (for ease of data set incorporation into novel
Unix, Linux, Free-BSD and Perl- or Shell-script design
and {\it vice-versa}), such a standardised nomenclature and
analytic treatment may provide {\it smart data sets} to advance
the solution. Far outperforming {\it smart programs} and {\it smart}
computational architecture, refinement in the definition itself 
of the structural, conformational and energetic phase space 
provides an allowance (place holder) in an analytic solution, for 
any observable generated by the operators sampling the
(wave)functions modeled.

The generation of accurate (wave)functions for the model 
becomes the focus of reductionist arguments, whereby the treatment 
of the electron density and correlation, thermodynamic and nuclear
terms, approximation levels and the mathematical description
employed to describe the physical space in relative proximity to
the nuclei (the {\it basis set}), must all be of sufficient accuracy
to account for all directing forces in folding.

The theoretical methods and level of theory employed in the 
modelling of polypeptides must therefore be calibrated to highly
deconvoluted and highly resolved experimental data
\cite{Alonso:Chass2004a,Alonso:Chass2002c,Alonso:Chin2004b,Alonso:Mirasol2004,Alonso:Chin2004a}.
In-depth analysis of experimental neutron and
x-ray diffraction (XRD), ultraviolet spectroscopy
(UV), mass-spectrometry (MS), electron-spin resonance (ESR), circular
dichroism (CD), infrared (IR), nuclear magnetic resonance (NMR)
experimental results
(among many others) must be incorporated in the derivation of an
accurate and tractable energetic partition function.

As each residue in the polypeptide chain has its own unique structural
and electronic contribution to the total evaluation, it may be prudent
to provide, at least as a {\it present} consideration and theoretical
exercise, the inclusion of subdivisions in the H (hydrophobic)
and P (polar) residue distinction proposed before.

For example, the Phenylalanine (Phe) residue is commonly treated as being
hydrophobic, yet it displays polar behaviour through interactions
of its aromatic sidechain ring with polarised atoms. Aromatic-amide
interactions are well established in the literature, among 
many other {\it weakly} polar intra-molecular ones
\cite{Alonso:Burkley1988,Alonso:Chipot1996,Alonso:McGaughey1998,Alonso:Tsuzuki2000,Alonso:Toth2001a,Alonso:Toth2001b,Alonso:Steiner2001,Alonso:Shi2002}.

A number of the stable Phe conformations, including the global minimum,
show such {\it weak} interactions to be important in. The peptide bond also 
shows differences to its {\it rigid and planar} description, with deviations
of $10-15^{o}$ being common, even for the global minumum
\cite{Alonso:Mirasol2004,Alonso:Chass2002b,Alonso:Chass2004b}. The impact
on the resonance structures of the peptide bond tautomer is not
trivial and results in large differences in the polarisation of the
consituent carbonyl ($C=O$) and amidic ($N-H$) elements. As these are
the atoms responsible for the majority of intra-molecular interactions
and hydrogen bonds, the peptide nitrogen's deviation from planarity
could become quite important in the folding process. Proline demands even
further scrutiny, as it shows a relatively large population of the cis-isomer
in proteins
\cite{Alonso:Eberhardt1996,Alonso:Milner-White1992,Alonso:Vitagliano2001}.

The H (hydrophobic) and P (polar) categories could be subdivided into glycyl
($^{g}H$), small ($^{s}H$), large ($^{l}H$), prolyl ($^{Pro}H$),
aromatic ($^{Ar}P$), oxygen- sulfur- and
nitrogen-containing ($^{o}P$, $^{s}P$ and $^{n}P$, respectively),
cationic ($^{+}P$) and anionic ($^{-}P$) subcategories.
The two last will be formed by protonation of His, Lys and Arg and
by deprotonation of Asp and Glu, respectively
(See Table \ref{Alonso:Categories}).

\begin{table}[t!]
\caption{Subdivision of H and P categories for naturally occurring amino acids}
\label{Alonso:Categories}
\begin{tabular}{rlrl}
\multicolumn{2}{c}{H} & \multicolumn{2}{c}{P} \\
\tableline
$^{g}H$ & Gly & $^{Ar}P$ & Tyr, Phe, Trp \\
$^{s}H$ &  Ala, Val & $^{o}P$ & Ser, Thr \\
$^{l}H$ & Leu, Ile & $^{s}P$ & Cys, Met \\
$^{Pro}H$ & Pro &  $^{n}P$ & Asn, Gln \\ 
 & & $^{+}P$ & His, Lys, Arg \\
 & & $^{-}P$ & Asp, Glu 
\end{tabular}
\end{table}

The quantum chemist or the computational molecular physicist
could then be periodically called upon to provide
the specified residues' structural, energetic, orbital and electron 
density results from his or her model peptides.

At the {\it present} time, we chemical, molecular, quantum and
theoretical physicists should be wary of any one model attempting
to fully describe the folding of polypeptides into biologically functional
proteins. All the while, approaching even our own works with
skepticism, demanding reproducibility and non {\it statistically-massaged}
proof.

The questions and challenges inherent in this problem require
the combined interpretation of results emerging from all
disciplines. A standardised definition and computed peptide
structure database \cite{Alonso:Chass2002a,Alonso:Sahai2003b}
must be established, able to be used by all
investigators from medicine and biology, through chemistry and
physics to computer science and informatics.
 
A race towards a full understanding, using solely one approach
will otherwise result in a large amount of repetition of works,
due to a lack of communication between differing disciplines.

The lure of simple models promising to incorporate all
elements and driving forces responsible for the folding of
polypeptides into biologically functional proteins, must be avoided
and the results interpreted with caution. Simple models will
certainly provide answers to a large number of questions,
however, they may also be devoid of finely tuned and highly
specific directing forces, with the models effectively {\it skewing
the fold} to a structure proximate in conformational hyperspace
to the native one, but with a reduction-in or absence-of 
biological activity.

\begin{quote}

We should be careful to get out of an experience only the 
wisdom that is in it --- and stop there; lest we be like the
cat that sits down on a hot stove-lid. She will never sit 
down on a hot stove-lid again; but also she will never sit
down on a cold one any more.

\begin{flushright}
--  Mark Twain  --
\end{flushright}

\end{quote}

A repeated, iterative refinement of the simple models
must therefore be undertaken, until the process suffers
from the law of diminishing returns whereby the difference
between the results of successively refined models, becomes
negligible.

The simple model that allows for {\it future} inclusion of 
more refined reductionist terms, even if only in practice,
will be afforded with a folding path to the true bases of
native structure and activity.

\section*{Epilogue}

We finish this tribute to Alberto Galindo quoting what 
the hungarian Nobel Laureate Albert Szent-Gy\"orgyi
said half a century ago and published in 1960
\cite{Alonso:Szent-Gyorgyi1960,Alonso:Csizmadia2004}:

\begin{quote}

{\it
The distance between those abstruse quantum mechanical calculations
and the patient bed may not be as great as believed.
}

\end{quote}


\vspace{20pt}

\noindent {\bf Aknowledgements:} We would like to thank M. Bueno,
P. Bruscolini, A. Cruz, E. Freire, V. Laliena, J. Onuchic, M. Orozco,
A. Rey, J. Sancho, D. H. Setiadi and our colleagues of the chemistry
section in the Zaragoza University for illuminating discussions.  This
work has been partially supported through the following research
contracts: grant \mbox{BFM2003-08532-C02-01}, MCYT (Spain) grant
\mbox{FPA2001-1813}, Grupo \mbox{Consolidado} Arag\'on Government
``Biocomputaci\'on y F\'{\i}sica de Sistemas Complejos'' and the
Global Institute Of COmputational Molecular and Materials Science
(GIOCOMMS, Toronto, Canada).  P. Echenique is supported by MECD
(Spain) FPU grant. I. G. Csizmadia wishes to thank the Hungarian
Ministry of Education for a Szent-Gy\"orgyi Visiting Professorship.

\end{document}